# High coherent frequency-entangled photons generated by parametric instability in active fiber ring cavity

Lei Gao[1,4], Hongqing Ran[1], Yulong Cao[1], Stefan Wabnitz[2,3], Zinan Xiao[1], Qiang Wu[1], Lingdi Kong[1], Ligang Huang[1], and Tao Zhu[1,4]

[1] *Key Laboratory of Optoelectronic Technology & Systems (Ministry of Education), Chongqing University, Chongqing 400044, China.*
[2] *Dipartimento di Ingegneria dell'Informazione, Elettronica e Telecomunicazioni, Sapienza Università di Roma, via Eudossiana 18, 00184 Roma, Italy.*
[3] *Novosibirsk State University, 1 Pirogova str, Novosibirsk 630090, Russia.*
[4] *corresponding authors: gaolei@cqu.edu.cn; zhutao@cqu.edu.cn*

High coherent frequency-entangled photons at telecom band are critical in quantum information protocols and quantum telecommunication. While photon pairs generated by spontaneous parametric down-conversion in nonlinear crystal or modulation instability in optical fiber exhibit random fluctuations, making the photons distinguishable among consecutive roundtrips. Here, we demonstrate a frequency-entangled photons based on parametric instability in an active fiber ring cavity, where periodic modulation of dispersion excites parametric resonance. The characteristic wave number in parametric instability is selected by the periodic modulation of resonator, and stable patterns with symmetric gains are formed. We find that the spectra of parametric instability sidebands possess a high degree of coherence, which is verified by the background-free autocorrelation of single-shot spectra. Two photon interference is performed by a fiber-based Mach-Zehnder interferometer without any stabilization. We obtain a Hong-Ou-Mandel interference visibility of 86.3% with a dip width of 4.3 mm. The correlation time measurement exhibits a linewidth of 68.36 MHz, indicating high coherence and indistinguishability among the photon pairs. Our results proves that the parametric instability in active fiber cavity is effective to generate high coherent frequency-entangled photon pairs, which would facilitate subsequent quantum applications.

As non-classical optical sources, quantum entangled photons are critical for testifying fundamental quantum mechanisms, such as quantum cryptography [1,2], quantum communications [3], quantum computing [4], violation of Bell inequalities [5]. Specially, frequency-entangled photons have been utilized to cancel dispersion in interferometers and to improve clock synchronization [6,7]. Spontaneous parametric down conversion in nonlinear bulk crystal and waveguide with high $\chi^2$ nonlinearity are frequently utilized for entangled photon pairs with high brightness [8-10]. However, challenges remain in the relative large material dispersion, where phase-matching conditions deteriorate for long interacting length, reducing conversion efficiency. Other drawbacks include the multi-mode characteristics, which require precise fiber-coupling alignment for quantum telecommunication.

The spontaneous four-wave-mixing process (FWM) based on $\chi^3$ nonlinearity is identified to be a promising candidate for producing reliable entangled photon pairs [11-15]. The modulation instability (MI) in optical fibers has been widely used to generate photon pairs with controllable frequency and polarization, by adjusting dispersion, nonlinearity, and spatial mode distribution. When the phase-matching conditions are satisfied, signal and idler photons are produced simultaneously at the expense of annihilation of two identical pump photons. Numerous schemes have been carried out based on single-mode fiber (SMF) [16], birefringence fiber [17], and photonic crystal fiber [18]. However, the spontaneous nonlinear process in MI-based schemes that stems from quantum noise seriously destabilize the photon pairs. Recent single-shot detections have found that the optical spectra produced by pumping pulses are totally different in consecutive round trips [19]. Thus, the generated photon pairs are distinguishable, making the subsequent quantum procedures non-deterministic. In addition, the extremely small correlation time between low coherent photons makes it difficult for multiphoton nonlocality and quantum teleportation. One example is the width of the Hong-Ou-Mandel (HOM) interference dip, which is determined by coherence time of the photon pairs [20]. In many experiments, optical filters with narrow bandwidth are utilized to increasing the coherence length of the photon pairs, resulting to a broader dip width [21]. Yet, the filtering do not optimize the randomness among each photon pairs, and it reduces the brightness of source.

The parametric instability (PI) produces gain sidebands with high coherence, which facilitates generation of deterministic photon pairs. As shown in Fig. 1(a), the PI is induced by parametric resonance due to periodic modulation of dispersion. Different from the MI, where symmetry breaking bifurcation favors the growth of modulation gains, the characteristic wave number in the PI is selected by the periodic modulation of resonator dispersion [22-27]. It is also known as the Faraday instability. This unique instability has been identified in fluids [28], Bose condensates [29], and nonlinear optics [30], where stable patterns are observed. When PI is excited in cavity, resonating can reduce the bandwidth of sidebands to a much smaller level [31-33]. While, the most intriguing property of PI in cavity-enhanced schemes are the fixed phases among consecutive photon pairs. Namely, the initial phase of each photon, $\varphi$, is determined by the resonating condition,

$$\varphi + nL = 2\pi m \qquad (1)$$

where, $m=1,2\ldots$, $n$ is the refractive index, $L$ is the cavity length. The identical initial phase makes the photon pairs generated in consecutive round trips are indistinguishable. This consistency highly facilitate the subsequent applications, such as quantum computation and quantum processing.

In this letter, we demonstrate the generation of high coherent frequency-entangled photons with small bandwidth based on PI in an active fiber ring cavity. It consists two sections of SMFs. One possesses normal dispersion, while the other is anomalous dispersed. Specifically, a section of erbium doped fiber (EDF) is optically pumped as the gain. When the PI arises, fundamental Stokes and anti-Stokes sidebands are excited. By performing the dispersive Fourier transformation detection (DFT), the corresponding single-shot spectra reveal that the PI sidebands are highly identical. Besides, we perform HOM interference detection based on a fiber-based Mach-Zehnder interferometer (MZI) without any stabilization feedback. The coincidence detection of the signal and ideal photons exhibits a HOM interference dip with a visibility larger than 86.3%. The correlation time measurement proves that the generated photon pairs exhibit high coherence and indistinguishability.

When we considering a fiber ring cavity with varying dispersion, the optical field $\psi$ evolution can be modeled by the nonlinear Schrödinger equation (NLSE) with varying Kerr nonlinearity $\gamma$ and second-order dispersion $\beta_2$ [22,23]

$$i\frac{\partial y}{\partial z} - \frac{b_2}{2}\frac{\partial^2 y}{\partial t^2} + g|y|^2 y + i\frac{a}{2}y = 0 \quad (2)$$

where, $\alpha$ denotes the linear loss (positive) or amplification (negative), respectively. We ignore the higher-order dispersion or Raman scattering, as they do not have a noticeable influence on the PI dynamics in our fiber cavity. Using the Floquet method transformation with the variable $U=\psi\alpha$, one obtains from (2)

$$i\frac{\partial U}{\partial z} - \frac{b_2}{2}\frac{\partial^2 U}{\partial t^2} + g'(z)|U|^2 U = 0 \quad (3)$$

Where, $\gamma'=\gamma-\alpha$. Supposing that $|u|^2 \ll P$, the perturbed continuous wave (CW) solution of the NLSE (3) is as following,

$$U(z,t) = \left[\sqrt{P} + u(Z,T)\right]\exp\{iPz\} \quad (4)$$

Therefore, the linearized equation for $u(Z,T)$ is obtained as

$$iu_Z = b^2 \frac{d(Z)}{2}u_{TT} + g'(Z)(u+u^*) \quad (5)$$

where, $Z=z/L_{nl}$, and $T=t/t_0$ (the nonlinear length $L_{nl}=1/(\gamma P)$, the dispersion length $L_d=t_0^2/|\beta_{2ave}|$, and the reference time unit $t_0=1$ ps). $\beta_2=L_{nl}/L_d$.

By writing the solution of (5) as the sum of Stokes and anti-Stokes sidebands, $u(Z,T)=a(Z)\exp\{i\Omega T\}+b(Z)\exp\{-i\Omega T\}$, we obtain two coupled linear ordinary differential equations with periodic coefficients for $a(Z)$ and $b(Z)$

$$a_Z = i\frac{b_2}{2}W^2 d(Z)a + ig'(Z)(a+b) \quad (6.1)$$

$$b_Z = -i\frac{b_2}{2}W^2 d(Z)b - ig'(Z)(a+b) \quad (6.2)$$

A linear stability analysis of equations (6) can be rigorously carried out numerically by the Floquet theory. Analytically, the PI sidebands appear around multiple frequencies

$$\omega_m = \sqrt{\frac{2}{\beta_{av}}\left[(\delta-2|u|^2) \pm \sqrt{(\frac{m\pi}{\Lambda})2+(|u|^2)^2}\right]} \quad (7)$$

where, $\delta$ is the cavity detuning, $\Lambda$ is the period, and $\beta_{ave}$ is the averaged dispersion. Namely, various orders of sidebands can be excited under different pump powers and cavity detunings [24,25].

The PI process in cavity-enhanced configuration permit the production of high coherent spectrum within each round trip. Therefore, we anticipate the excitation of frequency-entangled photon pairs with high consistency. To testify their non-classical behavior, a two photon interference measurement is performed. Figure 1(b) exhibits the HOM interference scheme, where the relative time delay of two paths is adjusted by a delay line (DL). After optical coupler 2 (OC2, 50:50), photons in two paths are filtered by two filters, then, coincidence counting (CC) is registered after simultaneous photons detections by two single photon detectors (SPDs). Therefore, the photons for the input ports of OC2 can be either $\omega_1$ or $\omega_2$. For such a setup, there are eight possible routines, and the state of output can be expressed as [34,35],

$$|\Psi>_{OUT} = |\Psi_{1,2}>_1 + |\Psi_{1,2}>_2 - |\Psi_{1,2}>_3 - |\Psi_{1,2}>_4 \quad (8)$$

When photon pair is indistinguishable, $|\Psi_{1,2}>_1 = |\Psi_{1,2}>_4$, then

$$|\Psi>_{OUT} = |\Psi_{1,2}>_2 - |\Psi_{1,2}>_3 \quad (9)$$

So coincidence probability of photon pair after interferometer

$$P_c \propto \int \frac{d\omega_1}{2\pi} \int \frac{d\omega_2}{2\pi} |<0|c(\omega_1)d(\omega_2)|\psi>_{IN}|^2 \quad (10.1)$$

$$c(\omega) = (\hat{a}e^{i\omega\tau} + \hat{b} + \hat{a} - \hat{b}e^{-i\omega\tau})/2 \quad (10.2)$$

$$d(\omega) = (\hat{a} + \hat{b}e^{i\omega\tau} - \hat{a}e^{-i\omega\tau} + \hat{b})/2 \quad (10.3)$$

$$|\psi>_{IN} = \frac{1}{2}\int\frac{d\omega_1}{2\pi}\int\frac{d\omega_2}{2\pi}A(\omega_1,\omega_2)\left[a^+(\omega_1)b^+(\omega_2)+a^+(\omega_2)b^+(\omega_1)\right]|0> \quad (10.4)$$

where, $A(\omega_1,\omega_2)$ is the biphoton spectral amplitude, $\tau$ is the relative delay time, $\sigma$ is the bandwidth of photon field [36].

If the difference of two optical paths was much larger than the coherent length of two photons, two photon interference will not occur between the photon pairs. The CC reads as a CW value. Nevertheless, when the optical path difference is within the coherence length, the photon wave packet overlaps in the OC2 [37,38]. There will be a maximum interference as far as the two-photon wave packet overlaps completely. The coincidence probability of photon pair after interferometer drop to zero. For a filter with Gaussian spectrum, the shape of the HOM dip can be expressed as [39],

$$f(\tau) \propto 1 - \frac{2VRT}{R^2+T^2}\exp(-\frac{\tau^2}{2\sigma^2}) \quad (11)$$

where, $V$, $R$, $T$, represent the visibility, the reflectance, the transmittance of OC, respectively.

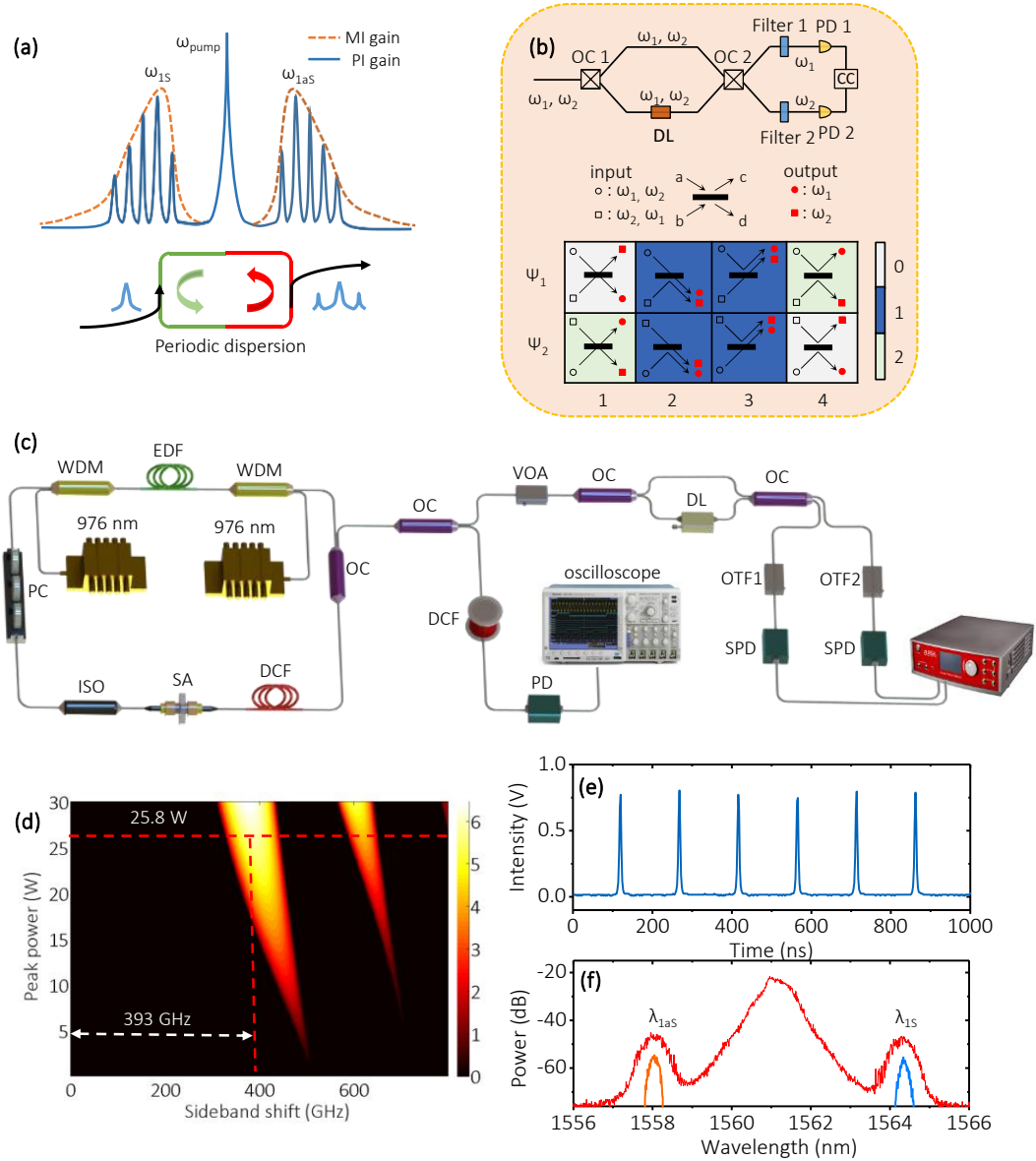

FIG. 1. Main experimental apparatus, and characteristics of PI sidebands. (a) Scheme of PI in resonator with periodic dispersion. (b) Scheme of HOM interference measurement. The different colors represent the number of detected photons. (c) Experimental setups for the frequency-entangled photon pairs based on PI in active fiber ring cavity, and the detail measurement methods. (d) Simulation results of PI sidebands. The dotted sideband shift for the peak power of 25.8 W is about 393 GHz. (e) Temporal train of pulses. (f) Optical spectrum of pulses, and two filtered wavelengths in the two channels.

The experimental system is schematically depicted in Fig.1(c). A ring fiber cavity contains 19.5 m DCF, 10.1m SMF, and 1 m EDF. It's pumped by two 976 nm CW lasers with power of 400 mW, through two wavelength division multiplexers (WDMs). A polarization independent optical isolator (ISO) is used for unidirection operation. The cavity detuning is optimized by a polarization controller (PC). The output is traced from an OC with a 10% output port. Additionally, a home-made saturable absorber (SA) is fabricated by filling reduced graphene oxide (rGO) flakes into cladding holes of a photonic crystal fiber [23,40]. The SA has a modulation depth of ~24%. The evanescent field interacting method permits only a very small percentage ($10^{-7}$) of light passing through the photonic crystal fiber interacting with rGO. Therefore, the thermal damage threshold can be increased substantially, which is inaccessible for conventional film-based SA. In the ring cavity, the dispersion of EDF, DCF, and SMF are 15.7, -38, and 18 ps/nm/km, respectively, corresponding to a net normal dispersion of 0.863 $ps^2$.

The optical spectrum is characterized by an optical spectrum analyzer (Yokogawa, AQ6370). Meanwhile, the corresponding single-shot spectra are recorded based on DFT, where periodic signals are stretched by a 500 m DCF with a dispersion of 150 $ps^2$ for frequency-to-time mapping, and subsequently fed to a 50 GHz photodetector connecting to a real time oscilloscope (Tektronix, DPO 71254) with a bandwidth of 12.5 GHz. Therefore, the optical resolution of the DFT is about 0.05 THz (0.2 nm in C band). After excitation of the PI, the laser intensity is attenuated by a variable optical attenuator (VOA), then the Stokes and anti-Stokes sidebands are filtered by two optical tunable filters (OTF, Santec, OTF-320), respectively. The full width at half maximum (FWHM) bandwidth of the two filters are 0.2 nm. For the HOM interference measurement, a fiber-based MZI is constructed by two OCs, in which a fiber-coupled optical DL is inserted to tune the path difference. Specifically, all fibers and devices are non-polarization maintaining and single-mode. The all-fiber configuration facilitate a convenient while robust connection for quantum telecommunication. Importantly, the fiber-based MZI is running at room temperature without any stabilization techniques. Namely, there is not stable phase difference due to the influence of environment. The photons in the two channels are detected by two identical InGaAs single-photon detectors under Geiger-mode within a high speed near-infrared single photon detection module (Auréa Technology), then the CC are registered. The two SPDs are triggered at a rate of 10 MHz, and the quantum efficiency is 20%. The gate window is 5 ns.

According to the experimental conditions, we follow the above parameters for simulations, Kerr nonlinearity $\gamma$=2 W$^{-1}$km$^{-1}$, a spatial period of the dispersion variation equal to the cavity length (30.6 m). The losses of the DCF, and the SMF are 0.5 dB/km, 0.2 dB/km, respectively. The gain of EDF compensates cavity loss exactly. An additional loss of 1dB is inserted for out-coupling loss and connector losses. The pulse duration is about 0.04 ns. The numerically computed PI sidebands for various peak powers are depicted in Fig. 1(d). Here, we focus on the fundamental sidebands, and neglect other high order sidebands as their energy is much lower.

By rotating the PC properly, PI can be excited with easiness. The temporal pulse train is shown in Fig. 1(e), where a period of 146 ns matches well with the cavity length. The averaged optical spectrum in Fig. 1(f) exhibits a center wavelength $\lambda_0$ at 1561.19 nm, and two sideband wavelengths, namely, $\lambda_{1AS}$=1558.048 nm, $\lambda_{1S}$=1564.338 nm. Therefore, the fundamental sideband spacing is 393 GHz, which is in accordance with the calculated value in Fig. 1(d) for an estimated intracavity peak power of 25.8 W. The corresponding consecutive single-shot spectra within 50 roundtrips in Fig. 2(a) clearly exhibit high consistency. The strong CW component is filtered automatically by the DFT. We observe well-fixed fine structures within the real-time optical spectra of the two sidebands, which are invisible in the averaged optical spectrum. It is totally different from the randomly evolving sidebands in MI in conventional SMF [19]. A frequency width of ~12 GHz (~0.1 nm in C band) is shown for the subpeaks within both the Stokes and anti-Stokes sidebands. It is interesting to remember that the resolution of the DFT detection is only 0.2 nm. Therefore, the obtained data maybe not accurate enough, yet, it denotes that the subpeaks formed due to PI within active fiber cavity possess narrow width, thus, high coherence.

The high coherence of the PI sidebands can be verified quantitatively by the background-free correlations of the fundamental anti-Stokes and Stokes sidebands as

$$\zeta = \langle S_n * S_n \rangle - \langle S_n \rangle * \langle S_n \rangle \qquad (12)$$

It present the average spectral autocorrelation of all single-shot spectra $S_n$ by subtracting the autocorrelation of the averaged spectrum. The coherence statistics of the two sidebands are depicted in Figs. 2(b) and 2(c). The positive central peak at zero frequency shift arises from the narrow lines in the single-shot spectra, whereas fluctuations at larger frequency shift denote cooperative (positive value) and competitive (negative value) interactions of the individual cavity modes [41]. Both autocorrelation analysis for the Stokes and anti-Stokes regions show quasi-periodicity along the frequency shift, indicating strong correlation among the single-shot spectra.

The PI sidebands with high coherence make it possible to generate frequency-entangled photons with high quality. To test the performance of the photon pairs, we attenuate the laser intensity, then filter the signal and idler channels. The inset in Fig. 3(a) give recorded photons in the two channels for a specific attenuation, where about 4.1 k photon pairs are produced per second. A small discrepance between the two channels originates from different losses, including splicing, insertion loose of delay line, detection efficiency of SPDs, and the transmission difference of OC. It can be optimized by adjustable attenuation.

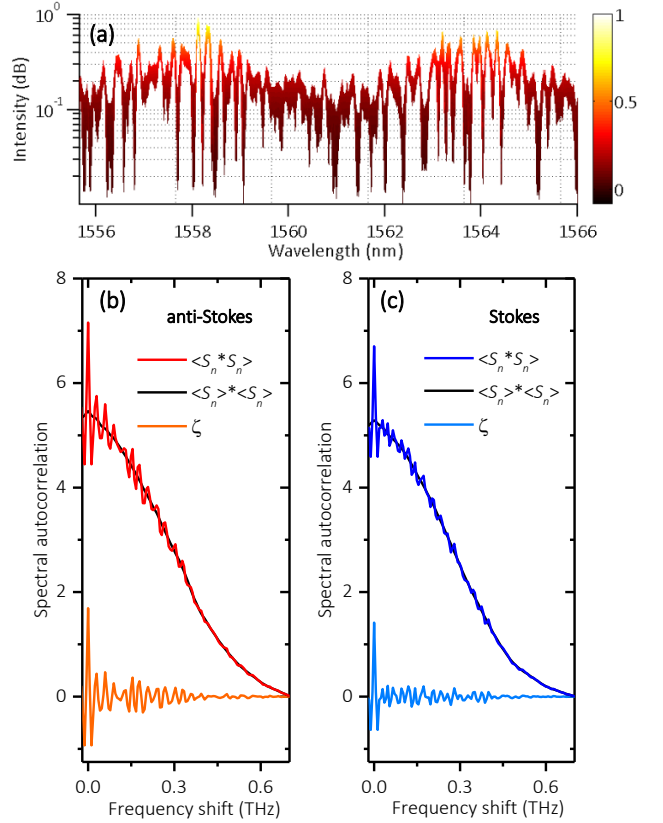

FIG. 2. Real time characteristics of PI sidebands. (a) 50 consecutive single-shot spectra of the fundamental Stokes and anti-Stokes bands. (b) Back-ground free autocorrelation of anti-Stokes (from 1555.6 nm to 1561.2 nm) and Stokes (from 1561.2 nm to 1566.8 nm) (c) computed from the single-shot spectra in (a).

The high coherence among conservative round trips are clearly revealed by the HOM interference measurement. According to the measurement setup, we scan the path difference, and plot the CC in Fig. 3(a). Each data point are averaged among 60 detections. When the two path difference approach to zero, two photon interference occurs. The maximum HOM interference results into a minimum CC. The experimental trace is fitted by the CC probability, 1-V$f(\tau)$. A net visibility of 86.3% is obtained by subtracting the accidental coincidences. This visibility can be further improved by optimizing the splitting ratio of OC, the polarization dependence of the optical devices, and most importantly, the path difference fluctuations due to temperature variations and wind vibrations in the laboratory. The fluctuations-induced distortion can be identified by the relative large standard deviations within the central region. The corresponding FWHM width is 4.3 mm, which is mainly determined by the filter bandwidth of 0.2 nm. This large bandwidth make more than one subpeaks account for the coincidence detection.

We perform a direct correlation time detection based on Hanbury Brown-Twiss configuration, where the detected signal of one SPD triggers the other. Figure 3(b) denotes the histogram of the CCs as a function of time delay between the two channels. After fitting with an exponential shape [31], a FWHM correlation time of 3.23 ns is obtained for the photon pair within one round trip, corresponding to a linewidth of 68.36 MHz. A similar correlation time of 3.14 ns is also obtained for the coincidence detection among two consecutive round trip. The normalized coincidence events within five consecutive photon pairs are also depicted in Fig. 3 (c), which exhibits high consistency. These results imply that the generated photon pairs based on PI possess high coherence and excellent indistinguishability.

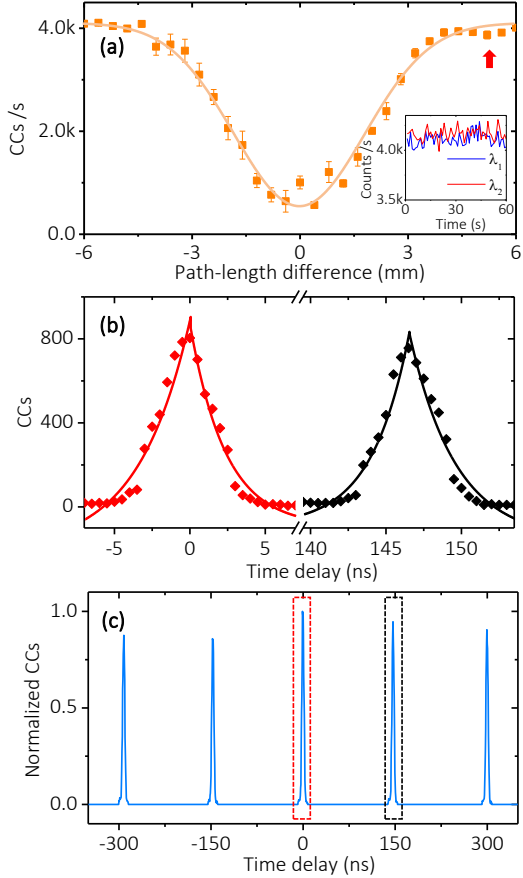

FIG. 3. (a) HOM measurement. CCs as a function of path-difference, superimposed by a Gaussian fitting curve. Each error bar corresponds to standard deviation of 60 detections. The inset is the photon counts for the two channels at a path difference of 5.1 mm. (b) Time correlation measurement for photons within the same and adjacent round trip. (c). Normalized CCs within five consecutive photon pairs.

Compared with the schemes based on MI in conservative system with CW or pulse pumping, this source can be highly brighter. Our scheme would generate as high as 6 million photon pairs/s, just determined by the cavity length and detection efficiency. Besides, the photon coherence is greatly enhanced by the PI process due to cavity configuration. The high coherent multiple subpeaks provide possibilities for wavelength division multiplexing to increase the bandwidth of modern quantum telecommunication. Moreover, our configuration do not require the period matching between extra pump laser and the cavity. There is not extra pulsed laser. The energy is provided by the amplified spontaneous emission generated by internal EDF. Finally, the PI in fiber resonator provide a flexible way to tune the frequency shift of sidebands by dispersion management. These virtues make the present scheme attractive as a simple, high efficient, and robust solution for reliable photon pairs.

In conclusion, we have developed and characterized a high coherent fiber-based frequency-entangled photon pairs. Periodic dispersion management in active fiber ring cavity excites PI, where Stokes and anti-Stokes sidebands are formed due to degenerate FWM. The single-shot spectra reveals that the sidebands possess subpeaks with narrow bandwidth, and high consistency among consecutive round trips leads to a high degree of coherence. We performed a fiber-based HOM interference measurement without any active stabilization. A HOM interference visibility of 86.3% is obtained, and the FWHM width of the dip is 4.3 mm. The correlation time measurement for the photon pairs exhibits a linewidth of 68.36 MHz, indicating high coherence and excellent indistinguishability. The results prove that the PI in active fiber cavity is attractive for the generation of high coherent frequency-entangled photon pairs. Our scheme also provide a simple, high efficient, and high brightness quantum source, which would facilitate subsequent applications.

This work was supported by the Natural Science Foundation of China (61635004, 61405023), the National Postdoctoral Program for Innovative Talents (BX201600200), and the National Science Fund for Distinguished Young Scholars (61825501). Stefan Wabnitz acknowledges support by the European Union's Horizon 2020 research and innovation program under a Marie Skłodowska-Curie program (691051), and the Russian Ministry of Science and Education (14.Y26.31.0017).

*References.*
1. N. Gisin, G. Ribordy, W. Tittel, and H. Zbinden, Rev. Mod. Phys. **74**, 145 (2002).
2. A. K. Ekert, J. G. Rarity, P. R. Tapster, and G. Massimo Palma, Phys. Rev. Lett. **69**, 1293 (1992).
3. V. Giovannetti, S. Lloyd, and L. Maccone, Science **306**, 1330 (2004).
4. H. Jeong, Phys. Rev. A **72**, 034305 (2005).
5. M. Kues, C. Reimer, P. Roztocki, L. R. Cortés, S. Sciara, B. Wetzel, Y. Zhang, A. Cino, S. T. Chu, B. E. Little, D. J. Moss, L. Caspani, J. Azaňa, R. Morandotti, Nature **546**, 622 (2017).
6. V. Giovannetti, S. Lloyd, and L. Maccone, Nature **412**, 417 (2001).
7. M. B. Nasr, B. E. A. Saleh, A. V. Sergienko, and M. C. Teich, Phys. Rev. Lett. **91**, 083601 (2003).
8. C. K. Hong, and L. Mandel, Phys. Rev. A **31**, 2409 (1985).
9. H. S. Eisenberg, G. Khoury, G. A. Durkin, C. Simon, and D. Bouwmeester, Phys. Rev. Lett. **93**, 193901 (2004).
10. J. W. MacLean, J. M. Donohue, and K. J. Resch, Phys. Rev. Lett. **120**, 053601 (2018).
11. X. Li, P. L.Voss, J. E. Sharping, and P. Kumar, Phys. Rev. Lett. **94**, 053601 (2005).
12. X. Bao, Y. Qian, J. Yang, H. Zhang, Z. B. Chen, T. Yang, J. W. Pan, Phys. Rev. Lett. **101**, 190501 (2009).
13. B. J. Smith, P. Mahou, O. Cohen, J. S. Lundeen, and I. A. Walmsley, Opt. Express **17**, 23589 (2009).
14. C. Reimer, M. Kues, P. Roztocki, B. Wetzel, F. Grazioso, B. E. Little, S. T. Chu, T. Johnston, Y. Bromberg, L. Caspani, D. J. Moss, and R. Morandotti, Science **351**, 1176 (2016).
15. J. A. Jaramillo-villegas, P. Imany, O. D. Odele, D. E. Leaird, Z. Y. Ou, M. Qi, and A. M. Weiner, Optica **4**, 655 (2017).
16. Y. M. Sua, J. Malowicki, M. Hirano, and K. F. Lee, Opt. Lett. **38**, 73 (2013).
17. K. G. Palmett, D. C. Delgado, F. D. Serna, E. O. Ricardo, J. M. Ruz, H. C. Ramirez, R. R. Alarcon, and A. B. U'Ren, Phys. Rev. A **93**, 033810 (2016).
18. L. Cui, X. Li, and N. Zhang, Phys. Rev. A **85**, 023825 (2012).
19. D. R. Solli, G. Herink, B. Jalali, and C. Ropers, Nat. Photon. **6**, 463 (2012).
20. C. K. Hong, Z. Y. Ou, and L. Mandel, Phys. Rev. Lett. **59**, 2044 (1987).
21. J. Liang, S. M. Hendrickson, and T. B. Pittman, Phys. Rev. A **83**, 033812 (2011).
22. C. Finot, F. Feng, Y. Chembo, and S. Wabnitz, Opt. Fiber Tech. **20**, 513 (2014).

23. L. Gao, T. Zhu, S. Wabnitz, M. Liu, and W. Huang, Sci. Rep. **6**, 24995 (2016).
24. M. Conforti, A. Mussot, A. Kudlinski, and S. Trillo, Opt. Lett. **39**, 4200 (2014).
25. F. Copie, M. Conforti, A. Kudlinski, A. Mussot, and S. Trillo, Phys. Rev. Lett. **116**, 143901 (2016).
26. M. Faraday, Phil. Trans. R. Soc. London **121**, 299 (1831).
27. T. B. Benjamin and F. Ursell, Proc. R. Soc. A **225**, 505 (1954).
28. P. Coullet, T. Frisch, and G. Sonnino, Phys. Rev. E **49**, 2087 (1994).
29. P. Engels, C. Atherton, and M. A. Hoefer, Phys. Rev. Lett. **98**, 095301 (2007).
30. A. M. Perego, N. Tarasov, D. V. Churkin, S. K. Turitsyn, and K. Staliunas, Phys. Rev. Lett. **116**, 028701 (2016).
31. Z. Y. Ou, and Y. J. Lu, Phys. Rev. Lett. **83**, 2556 (1999).
32. T. F. Langerfeld, H. M. Meyer, and M. Köhl, Phys. Rev. A **97**, 023822 (2018).
33. B. S. Shi, and A. Tomita, Phys. Rev. A **69**, 013803 (2004).
34. O. Cosme, and S. Padua. Phys. Rev. A **77**, 053822 (2008).
35. V. Giovannetti, L. Maccone, J. H. Shapiro, and F. N. C Wong, Phys. Rev. Lett. **88**, 183602 (2002).
36. T. S. Kim, H. O. Kim, J. H. Ko, and G. D. Park, J. Opt. Soc. Korea **7**, 113 (2003).
37. B. Yurke, S. L. McCall, J. R. Klauder, Phys. Rev. A **33**, 4033 (1986).
38. A. Trenti, M. Borghi, M. Mancinelli, H. M. Price, G. Fontana, and L. Pavesi, J. Opt. **18**, 085201 (2016).
39. H. Takesue, Appl. Phys. Lett. **90**, 204101 (2007).
40. L. Gao, T. Zhu, Y. J. Li, W. Huang, and M. Liu, IEEE Photon. Technol. Lett. **28**, 1245 (2016).
41. F. Intonti, V. Emiliani, C. Lienau, T. Elsaesser, V. Savona, E. Runge, R. Zimmermann, R. Nötzel, and K. H. Ploog, Phys. Rev. Lett. **87**, 076801 (2001).